\documentclass[superscriptaddress,aps,prl,showkeys,showpacs,twocolumn,10pt]{revtex4}
\usepackage[OT1,T1]{fontenc}
\usepackage{graphicx}
\usepackage{rotating}
\begin{document}

\title{ Comment on "Sequential Single-Pion Production Explaining the Dibaryon
  "$d^*(2380)$" Peak"}
\date{\today}

\newcommand*{\PITue}{Physikalisches Institut, Eberhard--Karls--Universit\"at
  T\"ubingen,  
Auf der Morgenstelle~14, 
72076 T\"ubingen, Germany}

\newcommand*{\York}{Department of Physics, University of York, Heslington,
  York, Y010 5DD, UK}

\author{M.~Bashkanov}   \affiliation{\York}
\author{H.~Clement}     \affiliation{\PITue}
\author{T.~Skorodko}     \affiliation{\PITue}

\begin{abstract}
In arxiv: 2102.05575  a two-step process $pn \to (pp) \pi^- \to (\Delta N) \pi^-
\to (d \pi^+) \pi^-$ was calculated by using experimental total cross sections
for  the single-pion production processes $pn \to pp \pi^-(I=0)$ and $pp \to d
\pi^+$. As a result the authors obtain a resonance-like structure for the
total $pn \to d\pi^+\pi^-$ cross section of about the right size and width of
the observed $d^*(2380)$ peak at an energy about 40 MeV below the $d^*(2380)$
mass. We object both the results of the sequential process calculation and its
presentation as an alternative to the dibaryon interpretation.
\end{abstract}


\maketitle

In what is called an unavoidable result of the sequential process the authors
of Ref.~\cite{seq} present their ansatz as an alternative to the dibaryon
interpretation of the resonance structure  $d^*(2380)$ with $I(J^P) = 0(3^+)$
observed in the 
$d\pi^+\pi^-$ channel \cite{d+-}. Note that this resonance has been observed
in all possible hadronic decay channels \cite{d00,np00,np+-,pp0-,np,npfull}
with the exception of the isoscalar $NN\pi$ channel. Its nonappearance in the
latter channel is in favor of a $d^*(2380)$ decay via an intermediate
$\Delta\Delta$ configuration as discussed in detail in
Ref.~\cite{isoNNpi}.   
A signal of $d^*(2380)$ has also been reported in photoexcitation of the
deuteron \cite{Ikeda,MB1,MB2,Ishikawa1,Ishikawa2,MAMI}. For recent reviews see
Refs. \cite{hcl,CPC}. 

We object both the results of the sequential process calculation and its
presentation as an alternative to the dibaryon interpretation as follows:

(i) For the isoscalar $pn(I=0) \to pp\pi^-$ reaction the WASA@COSY results
\cite{isoNNpi} are used in Ref.~\cite{seq} by enlarging the relatively small
statistical uncertainties enourmeously by adding in
quadrature a large systematic error due isospin violation. We think that since
the latter is not fluctuating randomly from energy point to energy point, it
does not behave like statistical uncertainties and can not just be added to
them. Only by this procedure the authors of Ref.~\cite{seq} can arrive at a
width as narrow as 70 MeV in a Breit-Wigner fit to the observed structure in
the isoscalar $pn \to pp\pi^-$ cross section. In our fits to this structure we
obtain a width of about 150 MeV, see Ref.s~\cite{isoNNpi,NN*}. Note that the
width of this first-step in the proposed sequential process is crucial for the
calculated narrow width of the structure in the final $d\pi^+\pi^-$ channel. 

(ii) The fact that the peak calculated for the $d\pi^+\pi^-$ channel misses
the measured peak by about 40 MeV is associated in Ref.~\cite{seq} with a
pretended experimental resolution of 20 MeV in $\sqrt s$. However, here the
authors of Ref. \cite{seq} mix the experimental resolution up with
the bin width used for the presentation of differential distributions in
Ref. \cite{d+-}. Furthermore, a finite experimental energy resolution affects
the width of a resonance structure, but not its position. 
The binning used for the presentation of total cross
section was 10 MeV in Refs. \cite{d+-,d00} --- see Fig.~1 --- and the high
precision COSY beam 
had a resolution in the sub-MeV range. The position of the peak calculated in
Ref. \cite{seq} rather is determined by the data point at 2.33 GeV in the
isoscalar single-pion production crossn section \cite{isoNNpi}, which
constitutes just a statistically insignificant excursion. In
Ref.~\cite{isoNNpi} it was demonstrated that there is no peak structure in the
data around 2380 MeV. Hence the concept of Ref.~\cite{seq} has no possiblity
to correct this 
failure in missing the correct resonance mass, which was independently
established in different two-pion production experiments
\cite{d+-,d00,np00,np+-,pp0-} and in $np$ scattering \cite{np,npfull}

\begin{figure} 
\centering
\includegraphics[width=0.99\columnwidth]{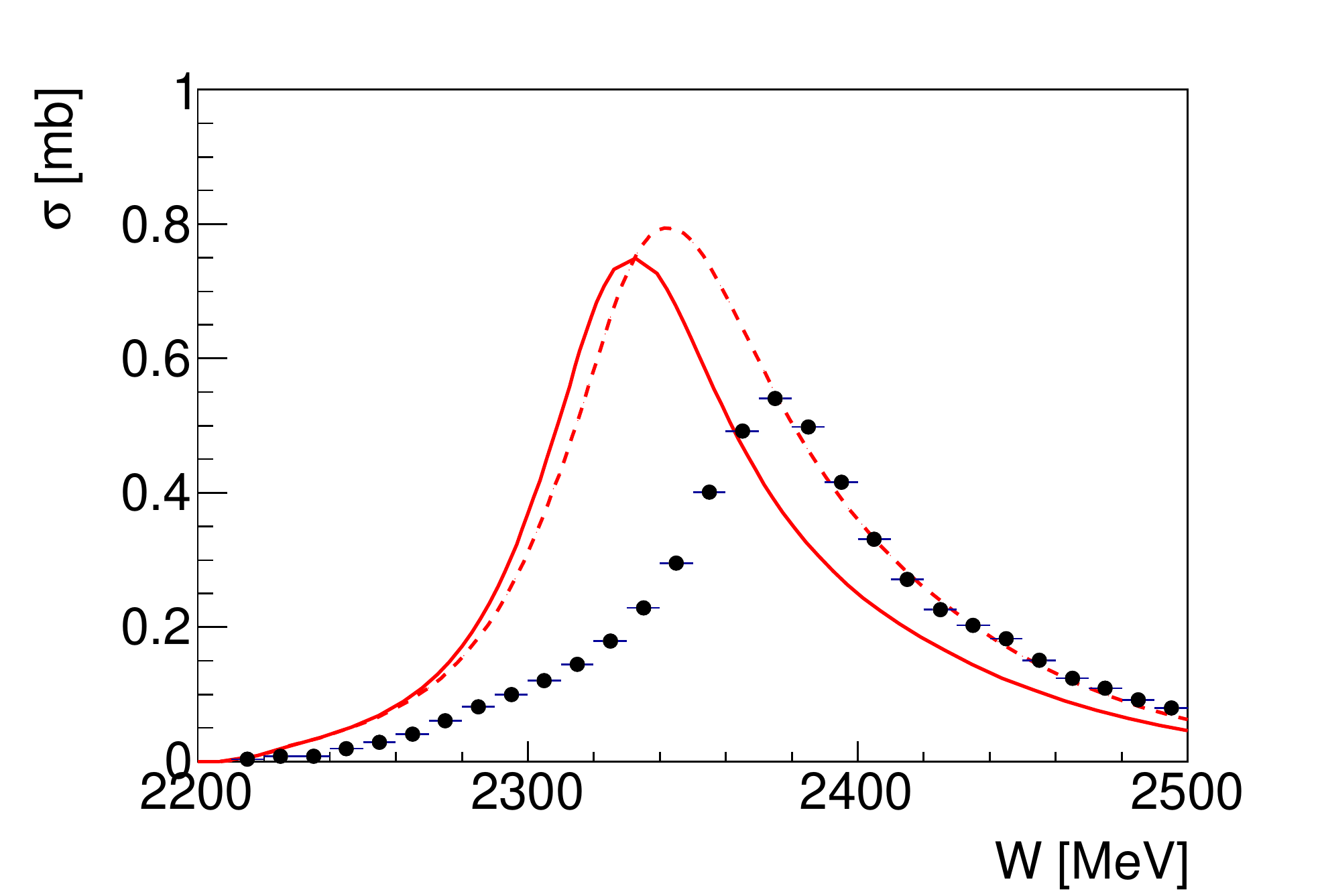}
\caption{\small (Color online) 
The isoscalar part of the total $pn \to d\pi^+\pi^-$ cross section in the
region of the dibaryon resonance $d^*(2380)$. Black filled dots represent the
experimental results from WASA-at-COSY \cite{d+-,d00}, the horizontal bars give
the binning width used. Red solid and dotted curves show the calculations of
Ref.~\cite{seq}. 
}
\label{fig1}
\end{figure}

(iii) As shown by many partial-wave analyses \cite{SAID,hcl} of the $pp \to
d\pi^+$ reaction the incident $^1D_2$ partial wave is the by far dominating
partial wave. Since in the isoscalar $N\pi$-invariant mass spectrum of the $np
\to pp\pi^-$ reaction the strength accumulates at highest masses (Fig. 6 of
Ref.\cite{isoNNpi}), it follows already from kinematics that the strength in the
associated $pp$-invariant mass spectrum accumulates at lowest masses --- in
accordance with a dominance of $S$- and $P$-waves between the final $pp$ pair
(see also the partial-wave analysis of Ref.~\cite{Sarantsev}). This is at 
variance with the requirements for incident $^1D_2$ $pp$-waves in the second
step reaction. With just low-energetic incident $S$- and $P$-waves available
the second step reaction provides only tiny cross sections and so does the full
sequential process with cross sections at least an order of magnitude smaller
than pretended in Ref.~\cite{seq}.  

(iv) Consistent $d^*(2380)$ signals have also been observed in the
non-fusion channels $np\pi^0\pi^0$ \cite{np00}, $np\pi^+\pi^-$ \cite{np+-} and
in $pp\pi^0\pi^-$ \cite{pp0-}, which are impossible to be explained by the
sequential process ansatz due to the absence of the triangle singularity.

(v) The concept of Ref.~\cite{seq} does not allow for any detailed check with
data, since it does not provide any differential distributions.

(vi) The sequential process cannot reproduce the observed pole in the
$^3D_3-^3G_3$ $np$-partial waves at 2380 MeV \cite{np}. Since the sequential
process produces a variety of spin-parity combinations, it produces also poles
in several partial waves at 2.33 GeV simultaneously, which are not in accord
with partial-wave analyses \cite{SAID,np}.

(vii) In contrast to the claim in Ref.~\cite{seq} various evidences for
$d^*(2380)$ signals have been observed in $\gamma d \to d\pi^0\pi^0$
\cite{Ishikawa1,Ishikawa2,MAMI} and $\gamma d \to pn$ reactions
\cite{Ikeda,MB1,MB2}, see, {\it e.g.}, the discussion in Ref.~\cite{CPC}. 

(viii) In order to demonstrate the invalidity of the sequential single-pion
production ansatz by yet another example, let us consider instead of the
isoscalar part now the isovector part of the $np \to pp\pi^-$
reaction. In this case we deal with the two-pion production
process $np(I=1) \to d\pi^+\pi^-$. Since the isovector part of the 
$pp\pi^-$ channel is larger than its isoscalar part by roughly a factor of
four \cite{isoNNpi} at the energy of the $d^*(2380)$ peak, we
would expect the cross secton for the isovector part of the $d \pi^+\pi^-$
channel to be larger than its isoscalar part by just this factor at the
position of $d^*(2380)$. In reality its is smaller by a factor of ten
\cite{d+-,CPC} and the sequential single-pion production ansatz fails again
vastly.

We finally note that it is already the knowledge of the pole in the $^3D_3$
$np$-partial wave and its associated Argand circle \cite{np}, which makes the
resonance structure in the various $NN\pi\pi$ channels unavoidable
\cite{CPC}. 

We acknowledge valuable discussions with V. Baru, A. Gal, Ch. Hanhart,
E. Oset, M. Platonova, A. Sarantsev, I. I. Strakovsky and C. Wilkin on this
matter. This work has been supported by DFG (CL 214/3-3).

\end{document}